\documentclass[twocolumn,tighten,trackchanges]{aastex61}
\usepackage{natbib,color,amsmath}
\usepackage[normalem]{ulem}
\usepackage{epstopdf}
\usepackage{hyperref}
\usepackage{graphicx}
\definecolor{red}{rgb}{1.0,0.0,0.0}

\newcommand{\Mj}[1]{$M_\mathrm{Jup}$}

\usepackage[utf8]{inputenc}

\begin{document}
\title{An updated visual orbit of the directly-imaged exoplanet 51 Eridani \lowercase{b} and prospects for a dynamical mass measurement with Gaia}

\correspondingauthor{Robert J. De Rosa}
\email{rderosa@stanford.edu}

\author[0000-0002-4918-0247]{Robert J. De Rosa}
\affiliation{Kavli Institute for Particle Astrophysics and Cosmology, Stanford University, Stanford, CA 94305, USA}

\author[0000-0001-6975-9056]{Eric L. Nielsen}
\affiliation{Kavli Institute for Particle Astrophysics and Cosmology, Stanford University, Stanford, CA 94305, USA}

\author[0000-0003-0774-6502]{Jason J. Wang}
\altaffiliation{51 Pegasi b Fellow}
\affiliation{Department of Astronomy, California Institute of Technology, Pasadena, CA 91125, USA}

\author[0000-0001-5172-7902]{S. Mark Ammons}
\affiliation{Lawrence Livermore National Laboratory, Livermore, CA 94551, USA}

\author[0000-0002-5092-6464]{Gaspard Duch\^ene}
\affiliation{Department of Astronomy, University of California, Berkeley, CA 94720, USA}
\affiliation{Univ. Grenoble Alpes/CNRS, IPAG, F-38000 Grenoble, France}

\author[0000-0003-1212-7538]{Bruce Macintosh}
\affiliation{Kavli Institute for Particle Astrophysics and Cosmology, Stanford University, Stanford, CA 94305, USA}

\author[0000-0002-9350-4763]{Meiji M. Nguyen}
\affiliation{Department of Astronomy, University of California, Berkeley, CA 94720, USA}

\author[0000-0003-0029-0258]{Julien Rameau}
\affiliation{Univ. Grenoble Alpes/CNRS, IPAG, F-38000 Grenoble, France}
\affiliation{Institut de Recherche sur les Exoplan{\`e}tes, D{\'e}partement de Physique, Universit{\'e} de Montr{\'e}al, Montr{\'e}al QC, H3C 3J7, Canada}

\author[0000-0002-5407-2806]{Vanessa P. Bailey}
\affiliation{Jet Propulsion Laboratory, California Institute of Technology, Pasadena, CA 91109, USA}

\author[0000-0002-7129-3002]{Travis Barman}
\affiliation{Lunar and Planetary Laboratory, University of Arizona, Tucson AZ 85721, USA}

\author{Joanna Bulger}
\affiliation{Institute for Astronomy, University of Hawaii, 2680 Woodlawn Drive, Honolulu, HI 96822, USA}
\affiliation{Subaru Telescope, NAOJ, 650 North A{'o}hoku Place, Hilo, HI 96720, USA}

\author[0000-0001-6305-7272]{Jeffrey Chilcote}
\affiliation{Department of Physics, University of Notre Dame, 225 Nieuwland Science Hall, Notre Dame, IN, 46556, USA}

\author[0000-0003-0156-3019]{Tara Cotten}
\affiliation{Department of Physics and Astronomy, University of Georgia, Athens, GA 30602, USA}

\author{Rene Doyon}
\affiliation{Institut de Recherche sur les Exoplan{\`e}tes, D{\'e}partement de Physique, Universit{\'e} de Montr{\'e}al, Montr{\'e}al QC, H3C 3J7, Canada}

\author[0000-0002-0792-3719]{Thomas M. Esposito}
\affiliation{Department of Astronomy, University of California, Berkeley, CA 94720, USA}

\author[0000-0002-0176-8973]{Michael P. Fitzgerald}
\affiliation{Department of Physics \& Astronomy, University of California, Los Angeles, CA 90095, USA}

\author[0000-0002-7821-0695]{Katherine B. Follette}
\affiliation{Physics and Astronomy Department, Amherst College, 21 Merrill Science Drive, Amherst, MA 01002, USA}

\author[0000-0003-3978-9195]{Benjamin L. Gerard}
\affiliation{University of Victoria, 3800 Finnerty Rd, Victoria, BC, V8P 5C2, Canada}
\affiliation{National Research Council of Canada Herzberg, 5071 West Saanich Rd, Victoria, BC, V9E 2E7, Canada}

\author[0000-0002-4144-5116]{Stephen J. Goodsell}
\affiliation{Gemini Observatory, 670 N. A'ohoku Place, Hilo, HI 96720, USA}

\author{James R. Graham}
\affiliation{Department of Astronomy, University of California, Berkeley, CA 94720, USA}

\author[0000-0002-7162-8036]{Alexandra Z. Greenbaum}
\affiliation{Department of Astronomy, University of Michigan, Ann Arbor, MI 48109, USA}

\author[0000-0003-3726-5494]{Pascale Hibon}
\affiliation{European Southern Observatory, Alonso de Cordova 3107, Vitacura, Santiago, Chile}

\author[0000-0001-9994-2142]{Justin Hom}
\affiliation{School of Earth and Space Exploration, Arizona State University, PO Box 871404, Tempe, AZ 85287, USA}

\author[0000-0003-1498-6088]{Li-Wei Hung}
\affiliation{Natural Sounds and Night Skies Division, National Park Service, Fort Collins, CO 80525, USA}

\author{Patrick Ingraham}
\affiliation{Large Synoptic Survey Telescope, 950N Cherry Ave., Tucson, AZ 85719, USA}

\author{Paul Kalas}
\affiliation{Department of Astronomy, University of California, Berkeley, CA 94720, USA}
\affiliation{SETI Institute, Carl Sagan Center, 189 Bernardo Ave.,  Mountain View CA 94043, USA}

\author[0000-0002-9936-6285]{Quinn Konopacky}
\affiliation{Center for Astrophysics and Space Science, University of California San Diego, La Jolla, CA 92093, USA}

\author[0000-0001-7687-3965]{James E. Larkin}
\affiliation{Department of Physics \& Astronomy, University of California, Los Angeles, CA 90095, USA}

\author{J\'er\^ome Maire}
\affiliation{Center for Astrophysics and Space Science, University of California San Diego, La Jolla, CA 92093, USA}

\author[0000-0001-7016-7277]{Franck Marchis}
\affiliation{SETI Institute, Carl Sagan Center, 189 Bernardo Ave.,  Mountain View CA 94043, USA}

\author[0000-0002-5251-2943]{Mark S. Marley}
\affiliation{NASA Ames Research Center, MS 245-3, Mountain View, CA 94035, USA}

\author[0000-0002-4164-4182]{Christian Marois}
\affiliation{National Research Council of Canada Herzberg, 5071 West Saanich Rd, Victoria, BC, V9E 2E7, Canada}
\affiliation{University of Victoria, 3800 Finnerty Rd, Victoria, BC, V8P 5C2, Canada}

\author[0000-0003-3050-8203]{Stanimir Metchev}
\affiliation{Department of Physics and Astronomy, Centre for Planetary Science and Exploration, The University of Western Ontario, London, ON N6A 3K7, Canada}
\affiliation{Department of Physics and Astronomy, Stony Brook University, Stony Brook, NY 11794-3800, USA}

\author[0000-0001-6205-9233]{Maxwell A. Millar-Blanchaer}
\altaffiliation{NASA Hubble Fellow}
\affiliation{Jet Propulsion Laboratory, California Institute of Technology, Pasadena, CA 91109, USA}

\author[0000-0001-7130-7681]{Rebecca Oppenheimer}
\affiliation{Department of Astrophysics, American Museum of Natural History, New York, NY 10024, USA}

\author{David Palmer}
\affiliation{Lawrence Livermore National Laboratory, Livermore, CA 94551, USA}

\author{Jennifer Patience}
\affiliation{School of Earth and Space Exploration, Arizona State University, PO Box 871404, Tempe, AZ 85287, USA}

\author[0000-0002-3191-8151]{Marshall Perrin}
\affiliation{Space Telescope Science Institute, Baltimore, MD 21218, USA}

\author{Lisa Poyneer}
\affiliation{Lawrence Livermore National Laboratory, Livermore, CA 94551, USA}

\author{Laurent Pueyo}
\affiliation{Space Telescope Science Institute, Baltimore, MD 21218, USA}

\author[0000-0002-9246-5467]{Abhijith Rajan}
\affiliation{Space Telescope Science Institute, Baltimore, MD 21218, USA}

\author[0000-0002-9667-2244]{Fredrik T. Rantakyr\"o}
\affiliation{Gemini Observatory, Casilla 603, La Serena, Chile}

\author[0000-0003-1698-9696]{Bin Ren}
\affiliation{Department of Physics and Astronomy, Johns Hopkins University, Baltimore, MD 21218, USA}

\author[0000-0003-2233-4821]{Jean-Baptiste Ruffio}
\affiliation{Kavli Institute for Particle Astrophysics and Cosmology, Stanford University, Stanford, CA 94305, USA}

\author[0000-0002-8711-7206]{Dmitry Savransky}
\affiliation{Sibley School of Mechanical and Aerospace Engineering, Cornell University, Ithaca, NY 14853, USA}

\author{Adam C. Schneider}
\affiliation{School of Earth and Space Exploration, Arizona State University, PO Box 871404, Tempe, AZ 85287, USA}

\author[0000-0003-1251-4124]{Anand Sivaramakrishnan}
\affiliation{Space Telescope Science Institute, Baltimore, MD 21218, USA}

\author[0000-0002-5815-7372]{Inseok Song}
\affiliation{Department of Physics and Astronomy, University of Georgia, Athens, GA 30602, USA}

\author[0000-0003-2753-2819]{Remi Soummer}
\affiliation{Space Telescope Science Institute, Baltimore, MD 21218, USA}

\author[0000-0002-5917-6524]{Melisa Tallis}
\affiliation{Kavli Institute for Particle Astrophysics and Cosmology, Stanford University, Stanford, CA 94305, USA}

\author{Sandrine Thomas}
\affiliation{Large Synoptic Survey Telescope, 950N Cherry Ave., Tucson, AZ 85719, USA}

\author[0000-0001-5299-6899]{J. Kent Wallace}
\affiliation{Jet Propulsion Laboratory, California Institute of Technology, Pasadena, CA 91109, USA}

\author[0000-0002-4479-8291]{Kimberly Ward-Duong}
\affiliation{Physics and Astronomy Department, Amherst College, 21 Merrill Science Drive, Amherst, MA 01002, USA}

\author{Sloane Wiktorowicz}
\affiliation{Department of Astronomy, UC Santa Cruz, 1156 High St., Santa Cruz, CA 95064, USA }

\author[0000-0002-9977-8255]{Schuyler Wolff}
\affiliation{Leiden Observatory, Leiden University, 2300 RA Leiden, The Netherlands}


\begin{abstract}
We present a revision to the visual orbit of the young, directly-imaged exoplanet 51 Eridani b using four years of observations with the Gemini Planet Imager. The relative astrometry is consistent with an eccentric ($e=0.53_{-0.13}^{+0.09}$) orbit at an intermediate inclination ($i=136_{-11}^{+10}$\,deg), although circular orbits cannot be excluded due to the complex shape of the multidimensional posterior distribution. We find a semi-major axis of $11.1_{-1.3}^{+4.2}$\,au and a period of $28.1_{-4.9}^{+17.2}$\,yr, assuming a mass of 1.75\,M$_{\odot}$ for the host star. We find consistent values with a recent analysis of VLT/SPHERE data covering a similar baseline. We investigated the potential of using absolute astrometry of the host star to obtain a dynamical mass constraint for the planet. The astrometric acceleration of 51~Eri derived from a comparison of the {\it Hipparcos} and {\it Gaia} catalogues was found to be inconsistent at the 2--3$\sigma$ level with the predicted reflex motion induced by the orbiting planet. Potential sources of this inconsistency include a combination of random and systematic errors between the two astrometric catalogs or the signature of an additional companion within the system interior to current detection limits. We also explored the potential of using {\it Gaia} astrometry alone for a dynamical mass measurement of the planet by simulating {\it Gaia} measurements of the motion of the photocenter of the system over the course of the extended eight-year mission. We find that such a measurement is only possible ($>98$\% probability) given the most optimistic predictions for the {\it Gaia} scan astrometric uncertainties for bright stars, and a high mass for the planet ($\gtrsim3.6$\,M$_{\rm Jup}$).
\end{abstract}

\section{Introduction}
The combination of relative astrometry of young, directly imaged substellar companions and absolute astrometry of their host stars is a powerful tool for obtaining model-independent mass measurements of this interesting class of objects (e.g., \citealp{Calissendorff:2018ee,Snellen:2018gc,Brandt:2018uj}). At young ages the luminosities of these objects encodes information of their formation pathway (e.g., \citealp{Marley:2007bf}), but interpretation is complicated by the degeneracy between initial conditions and the mass of the objects. While measurements from ESA's {\it Gaia} satellite \citep{GaiaCollaboration:2016cu} will be used to discover thousands of planets via the astrometric reflex motion induced on the host star \citep{Perryman:2014jr}, the vast majority of these detections will be around old stars where the observable signature of the initial conditions is lost, and photometric and spectroscopic characterization via direct imaging will be challenging if not prohibitively expensive. The intersection of these two techniques is giant planets and brown dwarfs detected around young ($\lesssim$100\,Myr) and adolescent ($\lesssim$1\,Gyr) nearby ($<50$\,pc) stars. Their proximity increases the amplitude of the astrometric signal, allowing for a more precise mass measurement, and their youth allows for tight constraints on the bolometric luminosity (e.g., \citealp{Chilcote:2017fv}), as well as detailed atmospheric characterization (e.g., \citealp{Rajan:2017hq}).

51 Eridani (51 Eri) is an F0IV \citep{Abt:1995eo} member of the 24--26\,Myr \citep{Bell:2015gw,Nielsen:2016ct} $\beta$~Pictoris moving group \citep{Zuckerman:2001go}. The star is part of a wide hierarchical triple system with the M-dwarf binary GJ 3305 \citep{Feigelson:2006ie}, with a $\sim60$\,kyr orbital period. As a nearby, young star, 51 Eri was a prime target for direct imaging searches to identify wide-orbit self-luminous giant planets. Observations obtained with the Gemini Planet Imager (GPI; \citealp{Macintosh:2014js}) revealed a planetary-mass companion at a projected separation of 13\,au \citep{Macintosh:2015ew}. The mass of the planet derived from the observed luminosity is a strong function of the initial entropy of the planet after formation. Considering the extrema of plausible initial entropies, the planet has a mass of either 1--2\,M$_{\rm Jup}$ for a high-entropy ``hot start'' formation scenario, or 2--12\,M$_{\rm Jup}$ for a low-entropy ``cold start'' scenario \citep{Marley:2007bf,Fortney:2008ez}. A measurement of the mass of the planet through a combination of relative and absolute astrometry would break this degeneracy, informing theories of giant planet formation at wide separations.

In this paper we present a study of the orbital parameters of 51~Eri~b, and investigate whether a dynamical mass measurement or constraint can be made by combining relative astrometry from GPI with absolute astrometry from {\it Hipparcos} and {\it Gaia}. We describe our ground-based observations in Section~\ref{sec:obs} and present an updated visual orbit fit in Section~\ref{sec:orbits}. We use this fit to predict the astrometric signal induced by the orbiting planet on the host star and compare to measured values derived from a combination of the {\it Hipparcos} and {\it Gaia} catalogues in Section~\ref{sec:accl}. We conclude with a prediction of the feasibility of a dynamical mass measurement of the planet using {\it Gaia} scan astrometry in Section~\ref{sec:gaia}.

\section{Observations and Data Reduction}
\label{sec:obs}
\begin{deluxetable*}{cccccc|cccc}
\tablecaption{51 Eri Gemini/GPI observing log and associated KLIP parameters\label{tbl:log}}
\tablehead{\colhead{UT Date} & \colhead{Filter} & \colhead{$N_{\rm exp}$} & \colhead{$t_{\rm int}\times n_{\rm coadds}$} & \colhead{$\Sigma t$} & \colhead{$\Delta$ PA} & \colhead{$\lambda_{\rm min}$--$\lambda_{\rm max}$} & $n_{\lambda}$ & \colhead{$m$} & \colhead{$n_{\rm KL}$}\\
 & & & (sec.) & (min.) & (deg) & ($\micron$) & & (px) & \\}
\startdata
2014-12-18\tablenotemark{a} & $H$ & 38 & $59.6\times1$ & 37.8 & 23.8 & 1.508--1.781 & 35 & 2 & 50\\
2015-01-30\tablenotemark{a} & $J$ & 45 & $59.6\times1$ & 44.7 & 35.1 & 1.130--1.334 & 35 & 2 & 50\\
2015-01-31\tablenotemark{a} & $H$ & 63 & $59.6\times1$ & 62.6 & 36.5 & 1.509--1.779 & 35 & 2 & 50\\
2015-09-01\tablenotemark{b} & $H$ & 93 & $59.6\times1$ & 92.5 & 43.8 & 1.512--1.777 & 35 & 2 & 50\\
2015-11-06 & $K_1$ & 52 & $59.6\times1$ & 51.7 & 26.4 & 1.903--2.177 & 33 & 2.5 & 50\\
2015-12-18 & $K_2$ & 103 & $59.6\times1$ & 102.4 & 71.8 & 2.131--2.316 & 25 & 2.5 & 50\\
2015-12-20 & $H$   & 148 & $59.6\times1$ & 147.1 & 80.1 & 1.511--1.776 & 35 & 2 & 50\\
2016-01-28 & $K_1$ & 97 & $59.6\times1$ & 96.4 & 55.5 & 1.941--2.172 & 28 & 2.5 & 50\\
2016-09-18 & $H$   & 94 & $59.6\times1$ & 93.4 & 49.9 & 1.511--1.777 & 35 & 2 & 50\\
2016-09-21 & $J$   & 83 & $29.1\times2$ & 82.5 & 53.1 & 1.133--1.332 & 35 & 1.5 & 50\\
2016-12-17 & $J$   & 84 & $29.1\times2$ & 81.5 & 44.7 & 1.135--1.331 & 35 & 1.5 & 50\\
2017-11-11 & $H$   & 44 & $59.6\times1$ & 43.7 & 27.7 & 1.508--1.777 & 35 & 2 & 50\\
2018-11-20 & $H$   & 59 & $59.6\times1$ & 58.7 & 32.9 & 1.509--1.780 & 35 & 2 & 50\\
\enddata
\tablenotetext{a}{Re-reduction of observations presented in \citet{Macintosh:2015ew}}
\tablenotetext{a}{Re-reduction of observations presented in \citet{DeRosa:2015jl}}
\end{deluxetable*}
\begin{figure}
\includegraphics[width=0.90\columnwidth]{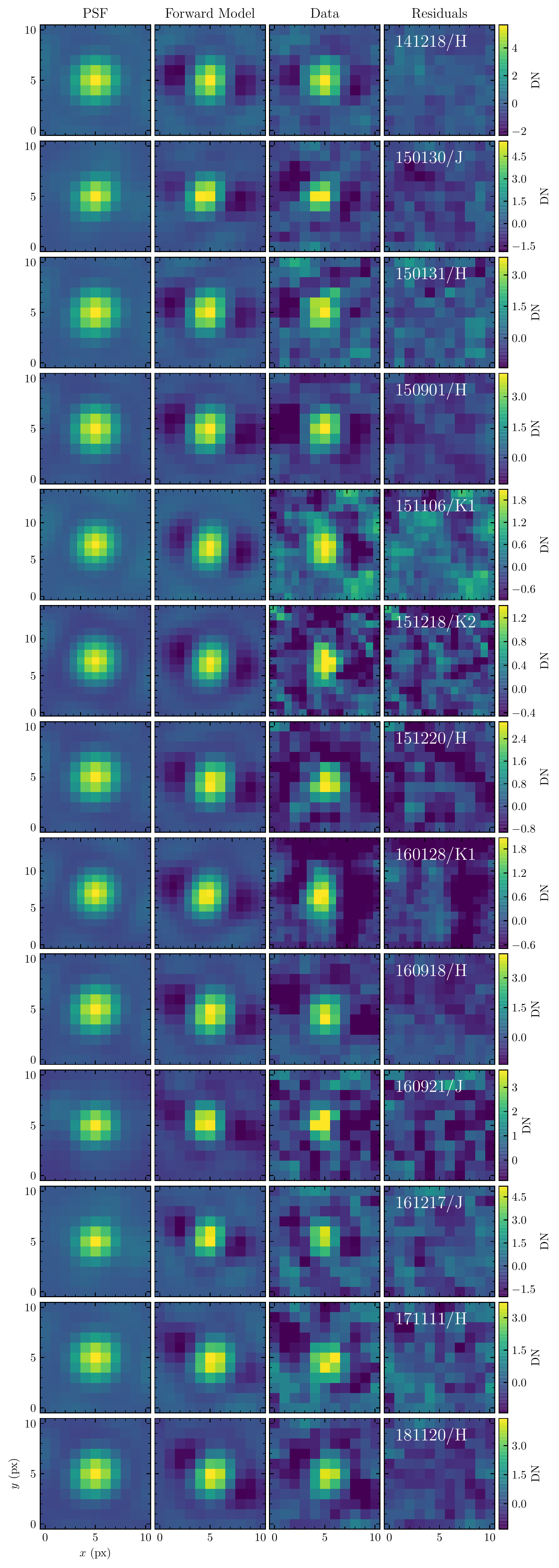}
\caption{GPI's PSF (first column), the BKA forward model (second column), the companion (third column), and residuals (fourth column) for each 51 Eri observation. The KLIP parameters used for each reduction are given in Table~\ref{tbl:log}. \label{fig:bka_stamps}}
\end{figure}

\subsection{Data acquisition and initial reduction}
51~Eri~b has been observed periodically with the Gemini Planet Imager (GPI; \citealp{Macintosh:2014js}) at Gemini South, Chile, during the Gemini Planet Imager Exoplanet Survey (GPIES; \citealp{Nielsen:2019td}) under program codes GS-2015B-Q-501 and GS-2017B-Q-501. GPI combines a high-order adaptive optics system and an apodized coronagraph to achieve high-contrast, diffraction-limited imaging over a $2\farcs8\times2\farcs8$ field-of-view. This field is then sent into an integral field unit that disperses the light at each point within the field-of-view into a low-resolution spectrum ($\lambda/\Delta\lambda$ between 35 at $Y$ to 80 at $K$). An observing log is given in Table \ref{tbl:log}; all observations were obtained in the default coronagraphic mode, but the filter and exposure time varied between epochs. All datasets were obtained in an Angular Differential Imaging (ADI, \citealp{Marois:2006df}) mode with the  Cassegrain rotator disabled causing the field of view to rotate in the instrument as the target transits overhead. Short observations of an argon lamp (30\,s) were obtained just prior to each science sequence to measure the positions of the microspectra in the raw frames which shift due to instrument flexure after large telescope slews. Observations of the arc were taken using the science filter except for sequences using the $K_1$ and $K_2$ filters where $H$ was used instead to minimize calibration overhead. Longer sets of observations of the argon lamp (300\,s) within each filter that are used for wavelength calibration, as well as darks of commonly-used exposure times, are obtained periodically at zenith according to the observatory's calibration plan.

Data were reduced using the GPI Data Reduction Pipeline (DRP v1.5; \citealp{Perrin:2014jh}), revision {\tt a494dd5}, as a part of the GPIES automated data processing architecture \citep{Wang:2018co}. Briefly, the DRP subtracts dark current, interpolates bad pixels using both a static bad pixel map and an outlier identification algorithm, constructs a 3-dimensional ($x$, $y$, $\lambda$) data cube, corrects for distortion over the field-of-view, and measures both the location and the flux of the four satellite spots (attenuated replicas of the central star generated via diffraction off a wire grid in the pupil plane) within each of the 37 wavelength slices of the final reduced data cube. The location of the central star behind the coronagraph was estimated from the location of these satellite spots. Observations previously published in \citet{Macintosh:2015ew} and \citet{DeRosa:2015jl} were reduced using an earlier version of the pipeline that contained several errors affecting the parallactic angle calculation \citep{2019arXiv191008659D}. These data were re-reduced using the updated version of the pipeline to ensure consistency.

\subsection{Point spread function subtraction}
The reduced data cubes were further processed using the Karhunen--Lo\`{e}ve Image Projection algorithm (KLIP; \citealp{Soummer:2012ig,Pueyo:2015cx}) to subtract the residual stellar halo that is not suppressed by the coronagraph, and the forward model-based Bayesian KLIP-FM Astrometry (BKA; \citealp{Wang:2016gl}) to measure the astrometry of the companion within each dataset. The forward model accounts for distortions in the instrumental PSF caused by the PSF subtraction process, providing a better match between the model used to fit the location of the companion. We used the implementation of KLIP and BKA available as a part of the {\tt pyKLIP} package\footnote{\url{http://bitbucket.org/pyKLIP} revision {\tt b3d97cd}} \citet{Wang:2015th}. Each wavelength slice of each data cube was high-pass filtered prior to PSF subtraction to remove low spatial frequency signals such as the residual seeing halo and instrumental background at $K$. An instrumental PSF was then constructed at each wavelength by averaging the four satellite spots in time. Wavelength channels with low throughput in the $K$-band filters were discarded where the satellite spots were too faint. The wavelength range ($\lambda_{\rm min}$--$\lambda_{\rm max}$) and number of wavelength channels ($n_{\lambda}$) used for each dataset are given in Table \ref{tbl:log}.

\begin{figure}
\includegraphics[width=1.0\columnwidth, trim=90 100 0 230]{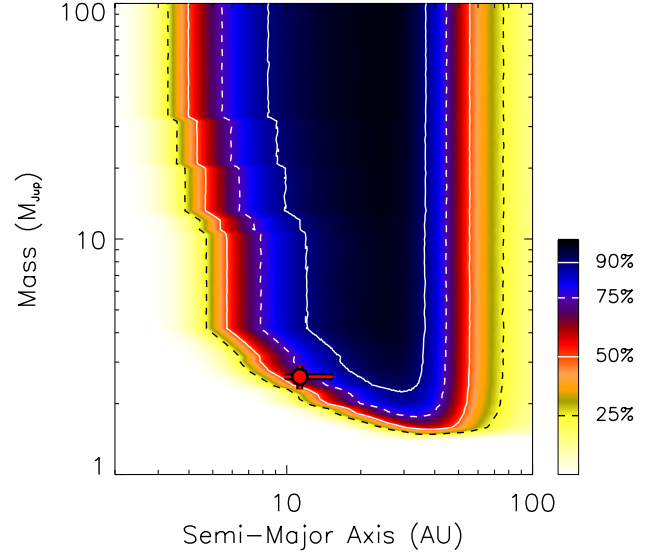}
\caption{Sensitivity to companions of 51~Eri as a function of their mass and semi-major axis. Contours denote 25\%, 50\%, 75\%, and 90\% sensitivity calculated after marginalizing over all other orbital elements. 51 Eri b is plotted, using the mass derived from the $H$-band luminosity, the \citet{Baraffe:2003bj} evolutionary models, and the \citet{Allard:2012fp} substellar atmosphere models. \label{fig:tongue}}
\end{figure}
KLIP PSF subtraction was performed within a single annulus centered on the star with a width of 16\,px at $J$ and $H$ and 20\,px at $K_1$ and $K_2$, and a radius such that the companion was centered between the inner and outer bounds of the annulus. The two main tunable parameters in the PSF subtraction process are the exclusion criteria $m$, defining the number of pixels an astrophysical source must move before an image can be included in the PSF reference library, and the number of Karhunen--Lo\`{e}ve modes $n_{\rm KL}$ used to reconstruct the stellar PSF. To explore the effects of the choice of these two parameters, we repeated the PSF subtraction using all combinations of $m \in \{1.0, 1.5, \dots, 4.0\}$ and $n_{\rm KL} \in \{1, 2, 3, 5, 10, 20, 50, 70\}$. The wavelength slices from each data cube after PSF subtraction were averaged resulting in one final PSF-subtracted image per epoch. We calculated point source sensitivity for each epoch and combined these into a single sensitivity map as a function of companion mass and semi-major axis using the algorithm described in \citet{Nielsen:2013jy,Nielsen:2019td}, shown in Figure~\ref{fig:tongue}.

\subsection{Relative astrometry}
The astrometry of the companion after each PSF subtraction of each epoch was then calculated using BKA. The forward model was created from the instrumental PSF given a specific combination of $m$ and $n_{\rm KL}$ and fit to the companion within the PSF-subtracted image within a small $11\times11$\,px box (or $15\times15$\,px at $K_1$ and $K_2$) centered on the estimated location of the companion. Posterior distributions for the position and flux of the companion and the correlation length scale \citep{Wang:2016gl} were sampled using the Markov-chain Monte Carlo (MCMC) affine-invariant sampler within the {\tt emcee} package \citep{ForemanMackey:2013io}. For each fit, 100 walkers were initialized near the estimated location for each parameter and were ran for 800 steps, with the first 200 discarded as burn-in. Uncertainties in the star centering (0.05\,px; \citealp{Wang:2014ki}) and the astrometric calibration (Table~\ref{tbl:astrometry}) from \citet{2019arXiv191008659D} were combined in quadrature with the statistical uncertainty derived from the MCMC posterior distributions.

The choice of KLIP parameters was driven by many factors: the location of the companion, the amount of field rotation, the spatial distribution of noise within the residual images (Fig. \ref{fig:bka_stamps}, fourth column), and the correlation (or lack thereof) between KLIP parameters and the measured astrometry. Large values for the exclusion parameter $m$ were preferred, although datasets with limited field rotation required a less restrictive setting. The parameters used for each dataset are given in Table \ref{tbl:log}, and the astrometry derived from the dataset processed with the selected parameters is given for each epoch in Table \ref{tbl:astrometry}.

\begin{deluxetable*}{ccccccccc}
\tablewidth{0pt}
\tablecaption{Relative astrometry of 51 Eri b using Bayesian KLIP Astrometry\label{tbl:astrometry}}
\tablehead{ \colhead{UT Date} & \colhead{MJD} & \colhead{Instrument} & \colhead{Filter} & \colhead{Plate scale} & \colhead{North offset} & \colhead{$\rho$} & \colhead{$\theta$} & \colhead{Reference} \\ 
& & & & (mas\,px$^{-1}$) & (deg) & (mas) & (deg) & }
\startdata
2014-12-18 & 57009.13 & Gemini/GPI & $H$ & $14.161\pm0.021$ & $0.17\pm0.14$ & $454.24\pm1.88$ & $171.22\pm0.23$ & 1 \\
2015-01-30 & 57052.06 & Gemini/GPI & $J$ & $14.161\pm0.021$ & $0.17\pm0.14$ &$451.81\pm2.06$ & $170.01\pm0.26$ & 1 \\
2015-01-31 & 57053.06 & Gemini/GPI & $H$ & $14.161\pm0.021$ & $0.17\pm0.14$ &$456.80\pm2.57$ & $170.19\pm0.30$ & 1 \\
2015-02-01 & 57054.25 & Keck/NIRC2 & $L^{\prime}$ & $9.952\pm0.002$ & $-0.252\pm0.009$ & $461.5\pm23.9$ & $170.4\pm3.0$ & 2 \\
2015-09-01 & 57266.41 & Gemini/GPI & $H$ & $14.161\pm0.021$ & $0.17\pm0.14$ & $455.10\pm2.23$ & $167.30\pm0.26$ & 1 \\
2015-11-06 & 57332.23 & Gemini/GPI & $K_1$ & $14.161\pm0.021$ & $0.21\pm0.23$ & $452.88\pm5.41$ & $166.12\pm0.57$ & 1 \\
2015-12-18 & 57374.19 & Gemini/GPI & $K_2$ & $14.161\pm0.021$ & $0.21\pm0.23$ & $455.91\pm6.23$ & $165.66\pm0.57$ & 1 \\
2015-12-20 & 57376.17 & Gemini/GPI & $H$ & $14.161\pm0.021$ & $0.21\pm0.23$ & $455.01\pm3.03$ & $165.69\pm0.43$ & 1 \\
2016-01-28 & 57415.05 & Gemini/GPI & $K_1$ & $14.161\pm0.021$ & $0.21\pm0.23$ & $454.46\pm6.03$ & $165.94\pm0.51$ & 1 \\
2016-09-18 & 57649.39 & Gemini/GPI & $H$ & $14.161\pm0.021$ & $0.32\pm0.15$ & $454.81\pm2.02$ & $161.80\pm0.26$ & 1 \\
2016-09-21 & 57652.38 & Gemini/GPI & $J$ & $14.161\pm0.021$ & $0.32\pm0.15$ & $451.43\pm2.67$ & $161.73\pm0.31$ & 1 \\
2016-12-17 & 57739.13 & Gemini/GPI & $J$ & $14.161\pm0.021$ & $0.32\pm0.15$ & $449.39\pm2.15$ & $160.06\pm0.27$ & 1 \\
2017-11-11 & 58068.26 & Gemini/GPI & $H$ & $14.161\pm0.021$ & $0.28\pm0.19$ & $447.54\pm3.02$ & $155.23\pm0.39$ & 1 \\
2018-11-20 & 58442.21 & Gemini/GPI & $H$ & $14.161\pm0.021$ & $0.45\pm0.11$ & $434.22\pm2.01$ & $149.64\pm0.23$ & 1 \\
\enddata
\tablerefs{(1) - this work; (2) - \citet{DeRosa:2015jl}.}
\end{deluxetable*}

\section{Updated Visual Orbit}
\label{sec:orbits}
\begin{deluxetable}{ccccc}
\tablewidth{0pt}
\tablecaption{Campbell elements and associated parameters describing the visual orbit of 51 Eridani b\label{tbl:elements}}
\tablehead{ \colhead{Parameter} & \colhead{Unit} & \colhead{Median $(\pm1\sigma$)} & \colhead{min. $\chi^2$ orbit} & \colhead{max. $\mathcal{P}$ orbit}}
\startdata
$P$ & yr & $28.1_{-4.9}^{+17.2}$ & 27.0 & 24.0\\
$a$ & $\arcsec$ &  $0.374_{-0.044}^{+0.140}$ & 0.363 & 0.338\\
$a$ & au & $11.1_{-1.3}^{+4.2}$ & 10.8 & 10.1\\
$r_{\rm peri}$ & au & $5.4_{-1.7}^{+3.8}$ & 4.7 & 3.9\\
$e$ & \nodata & $0.53_{-0.13}^{+0.09}$ & 0.57 & 0.61\\
$i$ & deg & $136_{-11}^{+10}$ & 138.9 & 144.5\\
$\omega$ & deg & $86_{-23}^{+23}$\tablenotemark{a} & 108.3 & 285.3\\
$\Omega$ & deg & $67_{-56}^{+63}$\tablenotemark{a} & 116.0 & 282.4\\
$\tau$ & $P$ & $0.56_{-0.22}^{+0.18}$ & 0.42 & 0.48\\
$T_0$ & MJD & $61735_{-712}^{+4824}$ & 61143 & 61202\\
$T_0$ & yr & $2027.9_{-2.0}^{+13.2}$ & 2026.3 & 2026.4\\
\enddata
\tablenotetext{a}{After wrapping $\Omega$ between 0--180\,deg}
\end{deluxetable}

\begin{figure*}
\includegraphics[width=1.0\textwidth]{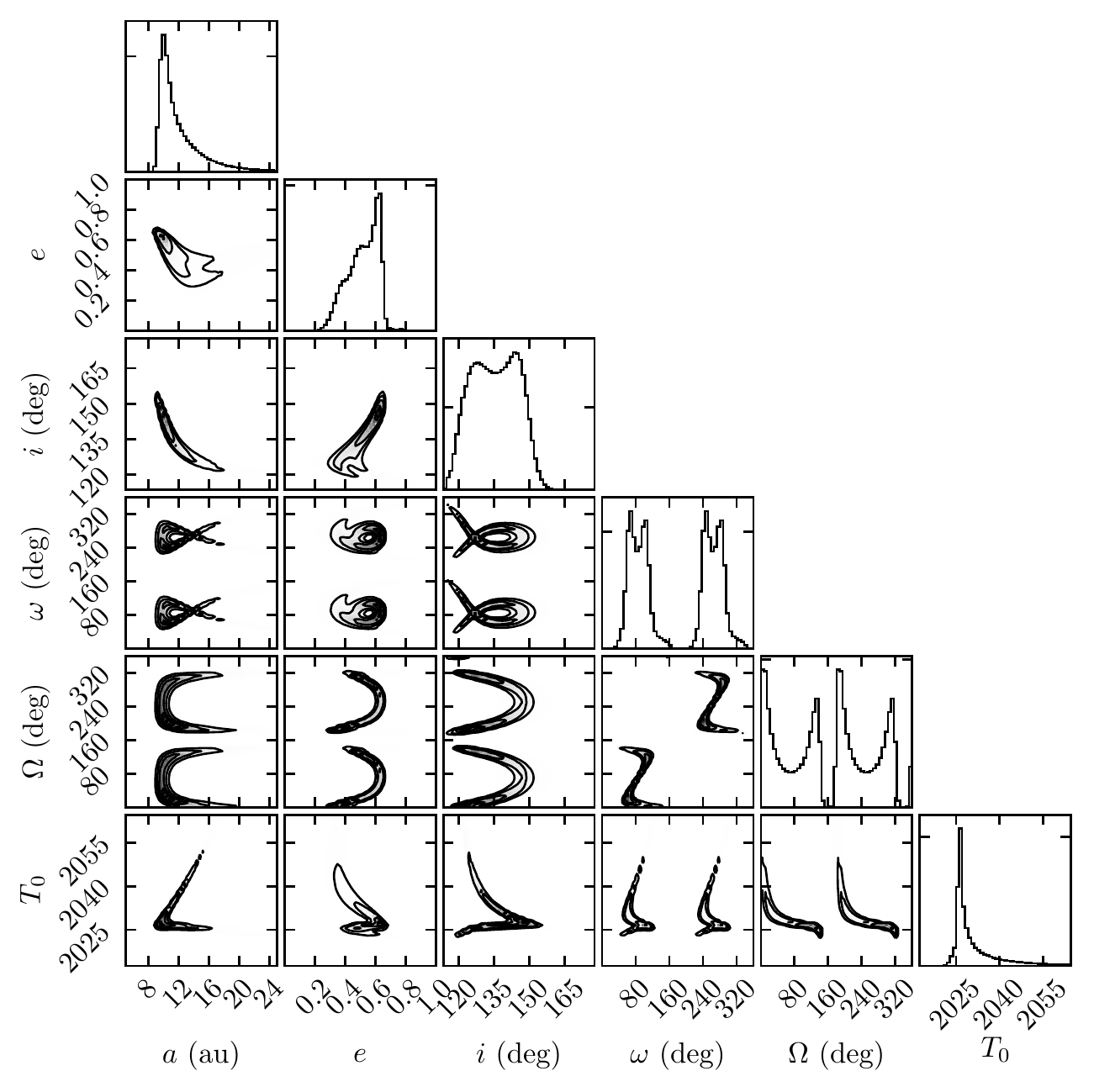}
\caption{Posterior distributions and their covariance for six of the orbital elements for the visual orbit of 51 Eri b.
\label{fig:corner}}
\end{figure*}
\begin{figure*}
\includegraphics[width=1.0\textwidth]{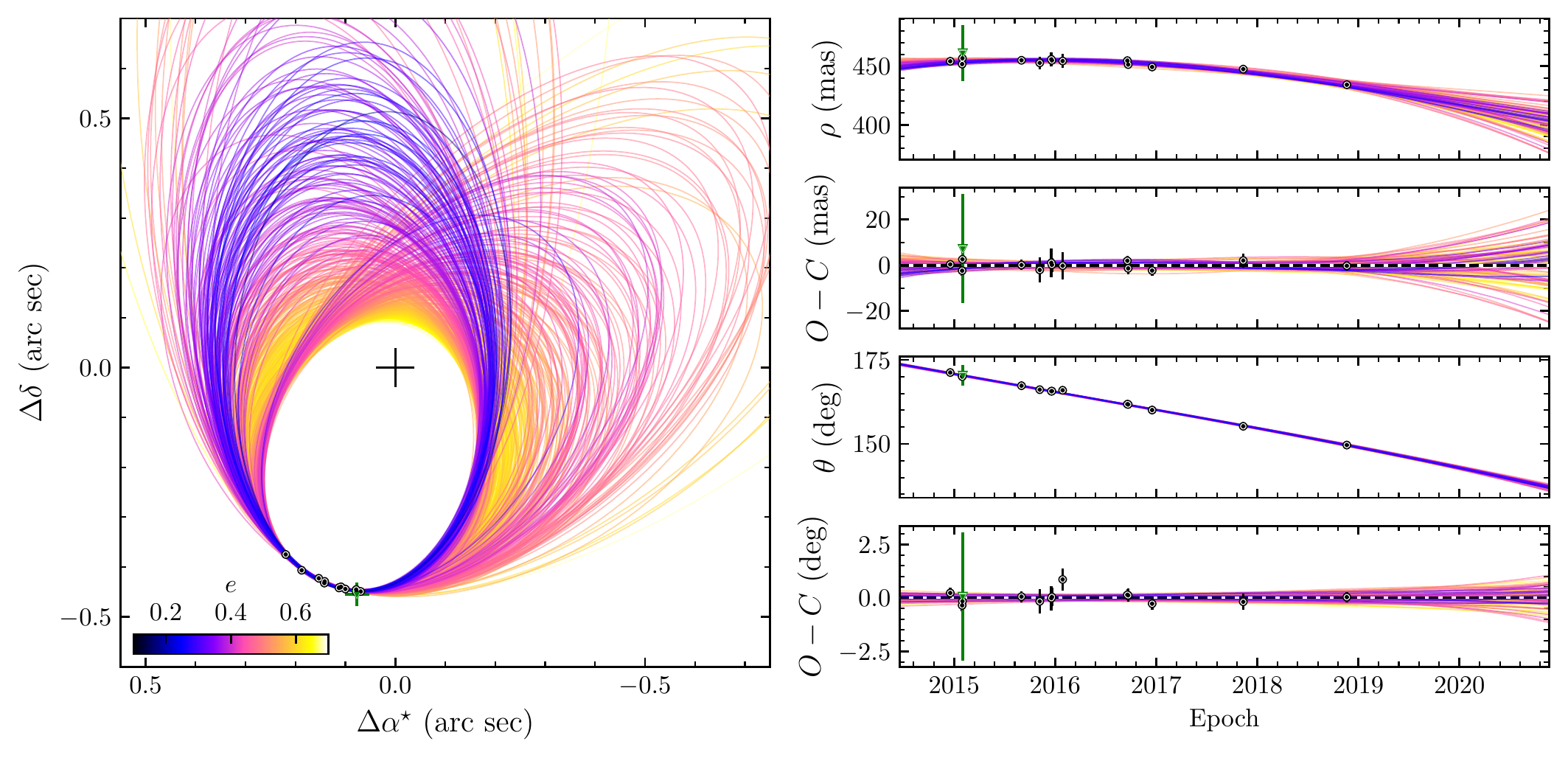}
\caption{(left) Five hundred visual orbits of 51 Eri b in the sky plane drawn from the MCMC chains, colored according to their eccentricity. Visual astrometry is overplotted from GPI (black) and NIRC2 (green). The location of the star is denoted by the cross. (right) Evolution of the separation (first row) and position angle (third row), and their associated residuals.
\label{fig:orbit}}
\end{figure*}

The relative astrometry presented in Table \ref{tbl:astrometry} was used to refine the orbital parameters of the planet. We used the parallel-tempered affine-invariant Markov chain Monte Carlo (MCMC) sampler within the {\tt emcee} package \citep{ForemanMackey:2013io} to sample the posterior distributions of six orbital elements (semi-major axis $a$, eccentricity $e$, inclination $i$, argument of periastron $\omega$, longitude of the ascending node $\Omega$, and epoch of periastron $\tau$), the parallax $\pi$, and the mass of the star $M_1$ and planet $M_2$. Rather than sampling $\omega$ and $\Omega$ individually, we sampled their sum ($\Omega+\omega$) and difference ($\Omega-\omega$) to speed up the convergence of the MCMC chains \citep{Beust:2014dj}. Standard priors on the orbital parameters were adopted; uniform in $\log a$, $e$, and $\cos i$. Gaussian priors were adopted on $\pi$ and $M_1$ based on the {\it Gaia} parallax measurement and uncertainties and literature estimates of the host star mass ($1.75\pm0.05$\,M$_{\odot}$; \citealp{Simon:2011ix}). Unlike systems where the period is constrained by the visual orbit (e.g., $\beta$~Pic; \citealp{Wang:2016gl}), we do not have sufficient coverage of the orbit to fit the total system mass directly and so we need to constrain the mass of the primary. We use a linear prior for $M_2$ between 1--15\,M$_{\rm Jup}$, encompassing the range of masses predicted from the measured luminosity and evolutionary models \citep{Macintosh:2015ew,Rajan:2017hq}. The visual orbit alone only constrains the total system mass; additional information (e.g., radial velocities, absolute astrometry) is required to constrain the mass ratio, and thus the masses of the two components.
 
We initialized 512 MCMC chains at each of 16 different temperatures (a total of 8192 chains). In the parallel-tempered framework the lowest temperature chains explore the posterior distributions of each parameter, while the highest temperature chains explore the priors. Each chain was advanced for $10^6$ steps and were decimated, saving the position of each walker every tenth step. The first tenth of the final decimated chains were discarded as a ``burn-in'' where the location of the walkers was still a function of their initial position. The trimmed and decimated chains yielded a total of 46,080,000 samples at the lowest temperature.

The posterior distributions for six of the orbital elements are shown in Figure~\ref{fig:orbit}, and are reported in Table~\ref{tbl:elements} along with the minimum $\chi^2$ and maximum probability (after accounting for the priors on the various parameters) orbits. We note that MCMC is not designed to find the minimum $\chi^2$, and it is likely that orbits with slightly lower $\chi^2$ could be found with a least-squares minimization algorithm using the best fit within the MCMC chains as a starting point. The quality of the fits to the astrometric record was typically good; the best fit orbit had $\chi^2=13.4$, corresponding to $\chi^2_{\nu} = 0.67$ assuming 20 degrees of freedom ($M_1$ and $M_2$ are dependent variables for a visual orbit fit), suggesting that the uncertainties on the astrometry were slightly overestimated. The visual orbit is plotted in Figure~\ref{fig:orbit} showing the predicted track of the planet in the sky plane, as well as the change in the separation and position angle of the planet as a function of time.

With the additional three years of astrometric monitoring we are beginning to constrain the eccentricity of the orbit of the planet. The fit presented in \citet{DeRosa:2015jl} only marginally constrained the eccentricity relative to the prior, only excluding the highest eccentricities. We similarly exclude high eccentricities $e>0.86$ is excluded at the 3$\sigma$ confidence level), but we also find that circular orbits are disfavored with the extended astrometric record. The preferred eccentricity is larger than for other directly imaged planets (e.g., \citealp{Wang:2018fd,Dupuy:2019cy}), although the sample size is currently too small to say whether it is unusually large. Interestingly---and most likely coincidentally---the median of the eccentricity distribution is consistent with the mean eccentricity of wide-orbit ($P>10^5$\,d) stellar companions to early-type (A6--F0) stars \citep{Abt:2005ke}.

We find a marginally smaller semi-major axis of $11.1_{-1.3}^{+4.2}$\,au with a significantly reduced uncertainty relative to \citet{DeRosa:2015jl}, and no significant change in the location and width of the inclination posterior distribution. There is a strong covariance between the eccentricity and inclination of the orbit, circular orbits are found closer to an edge-on configuration, while eccentric orbits are more face-on. A future radial velocity measurement of the planet has the potential to break this degeneracy well before continued astrometric monitoring is able to differentiate between the two families of orbits. In the context of additional undiscovered companions within the system, combining the semi-major axis and eccentricity distributions yields a periastron distance for the orbit of of $r_{\rm peri}=5.4_{-1.7}^{+3.8}$\,au. The posterior distribution on the mass of the planet is not constrained whatsoever relative to the uniform prior distribution described previously.

\subsection{Non-zero eccentricity}
The marginalized eccentricity posterior distribution shown in Figure~\ref{fig:corner} appears to suggest that circular ($e\sim0$) can be excluded at a high significance. This is in part due to the complex shape of the multidimensional posterior distribution. At small eccentricities the inclination is tightly constrained to $125\fdg2\pm0\fdg8$, and the longitude of the ascending node $\Omega$ is similarly constrained to one of two specific angles ($164\fdg 3\pm 0\fdg4$ and $344\fdg 3\pm 0\fdg 4$). At higher eccentricities these two parameters are far less constrained. As a consequence, the volume of phase space with allowable orbits with $e\sim0$ is considerably smaller than for more eccentric orbits despite the small difference in $\chi^2$, shifting the marginalized posterior distribution towards non-circular orbits.

To investigate whether or not we could exclude a circular orbit based on the current astrometric record we repeated the visual orbit fit described previously with the eccentricity and argument of periapse fixed at zero. We found a minimum $\chi^2$ of 18.7, corresponding to $\chi^2_{\nu}=0.85$ assuming 22 degrees of freedom. This is not significantly different from the best fit orbit found in the full fit described previously ($\chi^2_{\nu}=0.67$). Using the Bayesian information criterion, a circular orbit is preferred with a $\Delta {\rm BIC} = 1.4$, but not at a significant level. We therefore cannot reject the possibility that 51~Eri~b is on a circular orbit based on the current astrometric record, despite the shape of the marginalized posterior distribution shown in Figure~\ref{fig:corner}.

\subsection{Comparison with VLT/SPHERE astrometry}
Recently, \citet{Maire:2019vw} published a revision to the orbital parameters based on a combination of literature astrometry and three years of VLT/SPHERE observations of the system. The posterior distributions for the orbital elements are consistent between the two studies; both show that highly eccentric orbits are excluded by the current astrometric record. \cite{Maire:2019vw} note a potential systematic offset between the position angle measurements from GPI and SPHERE of $\theta_{\rm SPH} - \theta_{\rm GPI} = \Delta\theta = 1\fdg0\pm0\fdg2$ based on an independent reduction of the GPI data available in the archive. The source of such an offset can either be due to a systematic offset in the determination of the true north angle for both instruments, or an algorithmic issue caused by data reduction and/or post-processing.

We investigated this apparent discrepancy between the the two instruments by performing a joint fit to the astrometry presented in Table~\ref{tbl:astrometry} and \citet{Maire:2019vw} with two additional parameters; a multiplicative term to describe a relative magnification $\rho_{\rm SPH}/\rho_{\rm GPI} = \Delta\rho$, and an additive term to describe a constant position angle offset $\Delta\theta$ applied to the GPI measurements. The orbit fit was performed as previously, although the chains were thinned by a factor of 100 rather than 10. Compared with the joint fit performed by \citet{Maire:2019vw}, we find a more marginal offset between the two instruments, with a magnification of $\Delta\rho = 1.0050\pm0.0047$ and a position angle offset of $\Delta\theta = -0\fdg16\pm0\fdg26$. We do not see any significant offset between the GPI and SPHERE astrometric records using the astrometry presented in Table~\ref{tbl:astrometry} and \citet{Maire:2019vw}. This apparent discrepancy can be explained in part due to the revised astrometric calibration of GPI \citep{2019arXiv191008659D}, in which the north offset angle was changed by several tenths of a degree relative to the original calibration used by \citet{Maire:2019vw}. Repeating the orbit fit using the astrometry from Table~\ref{tbl:astrometry} but with the previous astrometric calibration yields a slightly different position angle offset of $\Delta\theta = 0\fdg28\pm0\fdg26$, significantly smaller than found by \citep{Maire:2019vw}. This suggests that the difference in the measured position angle offset could be algorithmic in nature, rather than a systematic calibration offset between the two instruments.

\section{Astrometric Acceleration}
\label{sec:accl}
\subsection{Absolute astrometry and inferred acceleration}
\begin{deluxetable}{lccc}
\tablewidth{0pt}
\tablecaption{{\it Hipparcos} and {\it Gaia} absolute astrometry of 51 Eri and inferred acceleration.\label{tbl:hip_gaia}}
\tablehead{\colhead{Property} & \colhead{Unit} & \colhead{Value}}
\startdata
\multicolumn{3}{c}{{\it Hipparcos} (1991.25)}\\
\hline
\multicolumn{3}{c}{HIP 21547}  \\
$\alpha$ & deg & $69.40044385\pm0.29$\,mas\tablenotemark{a} \\
$\delta$ & deg & $-2.47339207\pm0.20$\,mas  \\
$\mu_{\alpha^{\star}}$ & mas\,yr$^{-1}$ & $44.22\pm0.34$ \\
$\mu_{\delta}$ & mas\,yr$^{-1}$ & $-64.39\pm0.27$\\
$\pi$ & mas & $33.98\pm0.34$ \\
\hline
\multicolumn{3}{c}{{\it Gaia} (2015.5)}\\
\hline
\multicolumn{3}{c}{Gaia DR2 3205095125321700480}\\
$\alpha$ (cat.) & deg & $69.40074243852\pm0.1067$\,mas\tablenotemark{a} \\
$\alpha$ (corr.) & \nodata  & $69.40074243852\pm0.1163$\,mas\tablenotemark{a,b} \\
$\delta$ (cat.) & deg & $-2.47382451041\pm0.0724$\,mas \\
$\delta$ (corr.) & \nodata  & $-2.47382451041\pm0.0798$\,mas\tablenotemark{b} \\
$\mu_{\alpha^{\star}}$ (cat.) & mas\,yr$^{-1}$ & $44.352\pm0.227$  \\
$\mu_{\alpha^{\star}}$(corr.) & \nodata & $44.395\pm0.248$\tablenotemark{b} \\
$\mu_{\delta}$ (cat.) & mas\,yr$^{-1}$ & $-63.833\pm0.178$\\
$\mu_{\delta}$ (corr.) & \nodata & $-63.793\pm0.196$\tablenotemark{b} \\
$\pi$ (cat.) & mas & $33.5770\pm0.1354$  \\
$\pi$ (corr.) & \nodata & $33.5770\pm0.1477$\tablenotemark{b} \\
\hline
\multicolumn{3}{c}{Inferred proper motion difference}\\
\hline
$\mu_{\alpha^{\star}, {\rm G}}-\mu_{\alpha^{\star}, {\rm H}}$ & mas\,yr$^{-1}$ & $0.174\pm0.420$\\
$\mu_{\delta, {\rm G}}-\mu_{\delta, {\rm H}}$ & mas\,yr$^{-1}$ & $0.597\pm0.334$\\
$\mu_{\alpha^{\star}, {\rm H}}-\mu_{\alpha^{\star}, {\rm HG}}$ & mas\,yr$^{-1}$ & $-0.065\pm0.340$\\
$\mu_{\delta, {\rm H}}-\mu_{\delta, {\rm HG}}$ & mas\,yr$^{-1}$ & $-0.192\pm0.270$\\
$\mu_{\alpha^{\star}, {\rm G}}-\mu_{\alpha^{\star}, {\rm HG}}$ & mas\,yr$^{-1}$ & $0.110\pm0.249$\\
$\mu_{\delta, {\rm G}}-\mu_{\delta, {\rm HG}}$ & mas\,yr$^{-1}$ & $0.404\pm0.197$\\
\enddata
\tablenotetext{a}{Uncertainty in $\alpha^{\star} = \alpha \cos\delta$}
\tablenotetext{b}{After correcting for {\it Gaia} bright star reference frame rotation and the internal to external error ratio}
\end{deluxetable}

Astrometric measurements of 51 Eri were obtained from the re-reduction of the {\it Hipparcos} catalogue \citep{vanLeeuwen:2007dc} and the second {\it Gaia} data release (DR2; \citealp{GaiaCollaboration:2018io}), and are given in Table \ref{tbl:hip_gaia}. The {\it Gaia} catalogue is known to suffer from a number of systematics for bright stars like 51~Eri. The uncertainties in the position, proper motion, and parallax were inflated based on the ratio of internal to external uncertainties estimated by the {\it Gaia} consortium \citep{Arenou:2018dp}. The total uncertainty for each astrometric parameter was estimated using
\begin{equation}
    \sigma_{\rm ext} = \sqrt{k^2\sigma_i^2 + \sigma_s^2}
\end{equation}
where $\sigma_i$ is the catalogue uncertainty, $\sigma_s$ is a term representing the systematic uncertainty, and $k$ is a correction factor applied to the internal uncertainty. For bright stars ($G<13$), $k$ is assumed to be 1.08 and $\sigma_s$ is 0.016\,mas for position, 0.021\,mas for parallax, and 0.032\,mas\,yr$^{-1}$ for proper motion. Additionally, the bright star reference frame in {\it Gaia} DR2 was found to be rotating with respect to the stationary extra-galactic frame defined by distant quasars used for fainter stars. To correct for this, catalogue proper motions were rotated by the rotation matrix given in \citet{Lindegren:2018gy}, with the catalogue and rotation matrix uncertainties propagated using a Monte Carlo algorithm. Catalogue and corrected values for the {\it Gaia} DR2 astrometry are given in Table \ref{tbl:hip_gaia}; we exclusively used the corrected values for the analyses presented in this work.

We calculated three proper motion differentials from the two catalogues. The first ($\mu_{\rm G}-\mu_{\rm H}$) was calculated simply as the difference between the proper motion vector in the two catalogues ($\mu_{\rm H}$ for {\it Hipparcos}, and $\mu_{\rm G}$ for {\it Gaia}). Non-rectilinear and perspective effects that cause a change in the apparent motion of nearby stars of constant velocity are negligible at the distance of 51 Eri ($\lesssim1$\,$\mu$as\,yr$^{-1}$) and were therefore ignored. The two other differentials were calculated by comparing the instantaneous proper motion measured by each catalogue ($\mu_{\rm H}$, $\mu_{\rm G}$) to the proper motion derived from the absolute position of the star in both catalogues ($\mu_{\rm HG}$). Uncertainties were calculated using a Monte Carlo algorithm. The three proper motion differentials for 51 Eri are given in Table~\ref{tbl:astrometry}. A significant proper motion difference was measured in the declination direction for $\mu_{\rm G}-\mu_{\rm H}$ (1.8$\sigma$) and $\mu_{\rm G}-\mu_{\rm HG}$ (2.1$\sigma$); the other four values were not significantly different from zero. 

\subsection{Predicted acceleration due to 51 Eri b}
\label{sec:pred_accl}
\begin{figure}
\includegraphics[width=1.0\columnwidth]{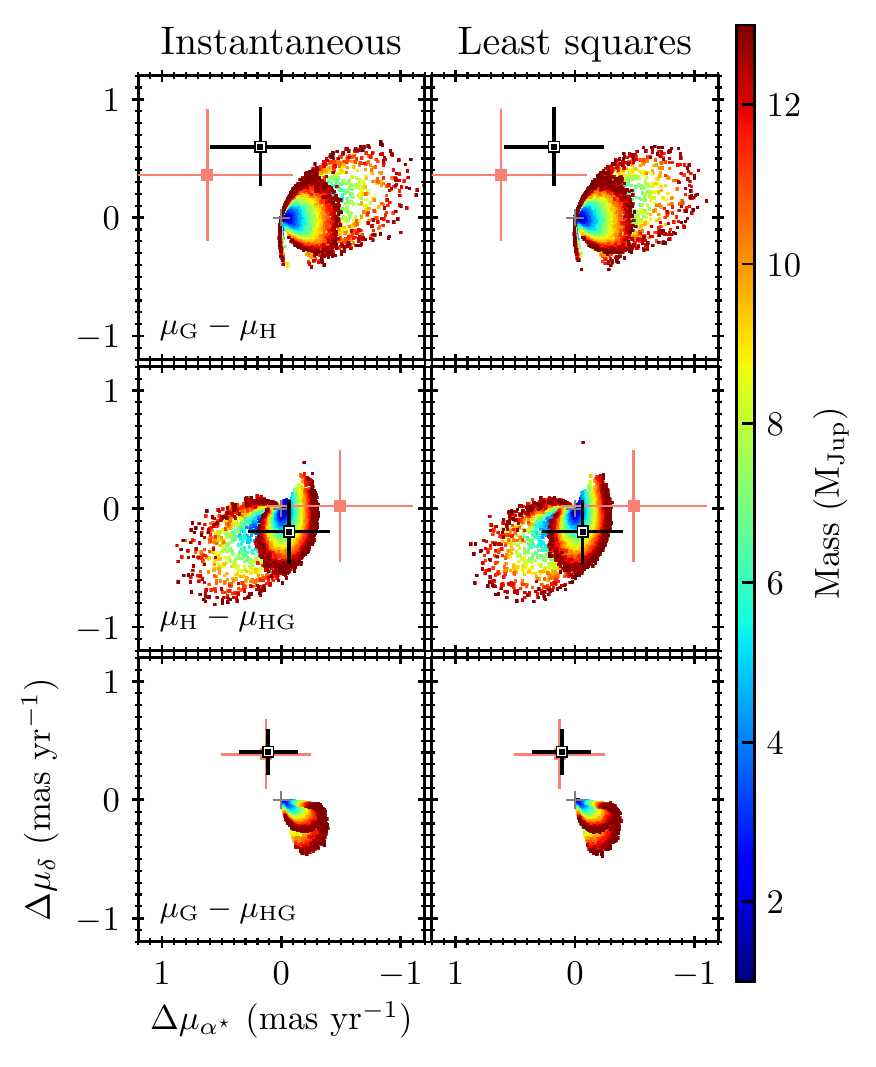}
\caption{Predicted astrometric signal induced on 51 Eri by the orbiting planet from the instantaneous proper motion of the photocenter (left column), and from the simplistic model of the {\it Hipparcos} and {\it Gaia} measurements (right column), both of which are calculated from the visual orbit fit. The color of the symbol denotes the mass of the planet for each visual orbit fit. For clarity, only 24000 fits drawn randomly from the posterior distributions are plotted. The accelerations in \ref{tbl:hip_gaia} are plotted (black squares) in addition to those computed by \citet{Brandt:2018dj} (red squares). \label{fig:accl1}}
\end{figure}
\begin{figure}
\includegraphics[width=1.0\columnwidth]{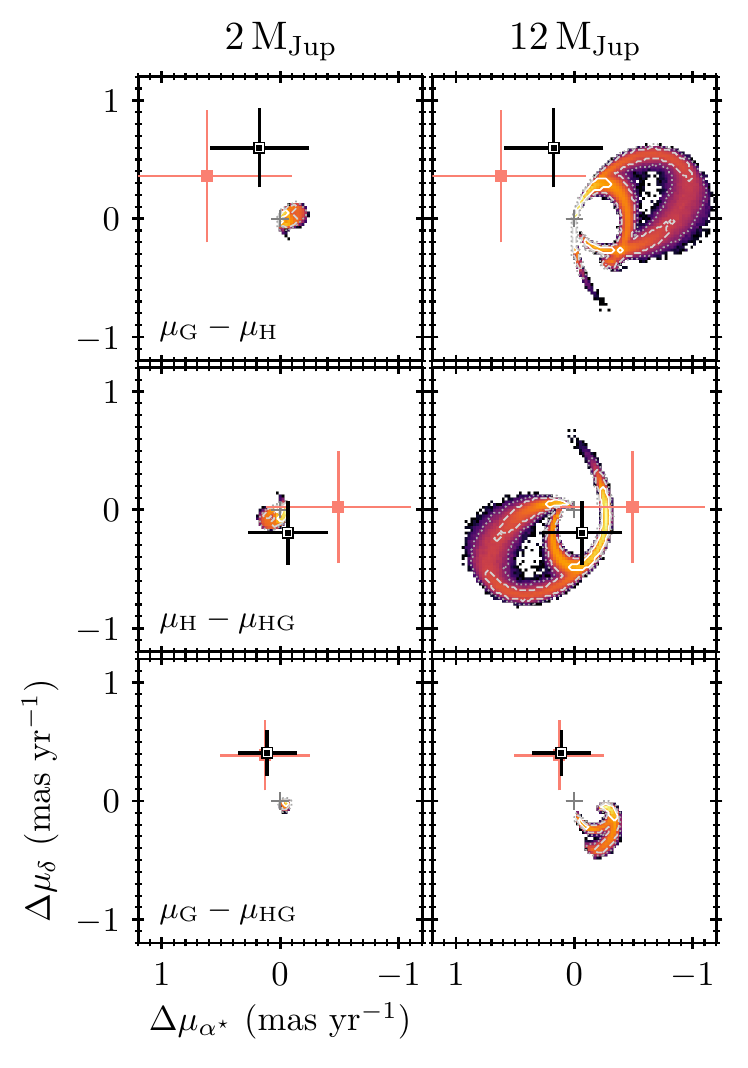}
\caption{Predicted astrometric signal induced on 51 Eri by the orbiting planet derived from the simulated {\it Hipparcos} and {\it Gaia} measurements for a 2\,M$_{\rm Jup}$ (left column) and 12\,M$_{\rm Jup}$ (right column) planet. Contours denote 1, 2, and 3$\sigma$ credible regions, the color scale is the logarithmic count of the orbits within each bin of the two-dimensional histogram. The measured signal is denoted by the black symbol, and the accelerations computed by \citet{Brandt:2018dj} are also shown (red squares) \label{fig:accl2}}
\end{figure}
We predicted the astrometric reflex motion induced by the orbiting planet on 51 Eri using the visual orbit fits described in Section~\ref{sec:orbits}. This signal was predicted using two different algorithms that produced consistent results. The first was based on the assumption that the {\it Hipparcos} and {\it Gaia} proper motion measurements were instantaneous. This assumption is likely valid for {\it Gaia} due to the current wide separation of the planet, but may not be valid for {\it Hipparcos} for more eccentric orbits. In this algorithm the instantaneous proper motion of the photocenter was calculated at the reference epoch for both missions ($\mu_{\rm H}$, $\mu_{\rm G}$). The long-term proper motion ($\mu_{\rm HG}$) was calculated as the difference in the photocenter position at both epochs divided by the 24.25\,yr baseline; a 10\,mas shift in the position of the photocenter would manifest itself as a change in the proper motion of the star of $\sim0.4$\,mas\,yr$^{-1}$. We assumed that the photocenter was centered on the host star; the planet contributes negligible flux within the {\it Hipparcos} and {\it Gaia} bandpasses.

The second algorithm was a simplistic simulation of the individual {\it Hipparcos} and {\it Gaia} measurements of the photocenter during the two missions. A simulated {\it Hipparcos} measurement was constructed by generating a one-dimensional abscissa measurement using a nominal set of astrometric parameters for the 51 Eri system barycenter. We adopted the {\it Hipparcos} catalogue values, but the results should not be sensitive to small changes in the reference position, parallax, and proper motion of the system barycenter. The abscissa was constructed using the procedure described in \citet{Sahlmann:2010hh}, and the scan epochs, angles, and parallax factors for 51 Eri provided in the {\it Hipparcos} Intermediate Astrometric Data (IAD) catalogue \citep{vanLeeuwen:2007du}. The abscissa was perturbed by the predicted photocenter orbit for a given sample within the MCMC chains. The offset between the photocenter and system barycenter at each epoch in the $\alpha^{\star}$ and $\delta$ directions were weighted by the scan angle of the satellite at that epoch.

Using this simulated abscissa measurement we predict what astrometric parameters would have been reported by {\it Hipparcos}. As the abscissa is a linear function of the five astrometric parameters ($\alpha^{\star}$, $\delta$, $\pi$, $\mu_{\alpha^{\star}}$, $\mu_{\delta}$), a unique solution could be found rapidly through a simple matrix inversion. This allowed us to compute the five astrometric parameters that would have been measured by {\it Hipparcos} for each of the $4\times10^7$ orbits described in Section~\ref{sec:orbits}. This process was repeated to simulate a {\it Gaia} measurement of the motion of the photocenter using the scan epochs, angles, and parallax factors predicted for 51 Eri using the {\it Gaia} Observing Schedule Tool\footnote{\url{https://gaia.esac.esa.int/gost/}}.

\subsection{Comparison with measured acceleration}
The predicted proper motion differentials calculated using these two algorithms are shown in Figure~\ref{fig:accl1}. The two algorithms are in excellent agreement, most likely due to the limited amount of curvature in the orbit of the photocenter during the {\it Hipparcos} and {\it Gaia} missions. The astrometric signal predicted using the second algorithm is plotted in Figure~\ref{fig:accl2} for orbits with a mass for the planet of $2.0\pm0.5$\,M$_{\rm Jup}$ and $12\pm0.5$\,M$_{\rm Jup}$, corresponding to the range of plausible masses for the planet based on evolutionary models, drawn from the visual orbit MCMC fit. It is evident that there is a significant discrepancy between the predicted proper motion differentials induced by the orbit of 51 Eri b and those measured with the {\it Hipparcos} and {\it Gaia} catalogue values. The measured differential between {\it Gaia} and the long-term proper motion ($\mu_{\rm G}-\mu_{\rm HG}$) is notably discrepant; the direction of this acceleration is in the opposite direction predicted from the visual orbit, and the 1$\sigma$ credible region for the predicted signal is significantly displaced from the measured value. A similar problem is seen for the difference between the two catalogue proper motions ($\mu_{\rm H}-\mu_{\rm G}$), although both the measurement uncertainties and the 1$\sigma$ credible interval of the predicted signal are larger. The two discrepant measurements both rely on the {\it Gaia} proper motion; the measured $\mu_{\rm H}-\mu_{\rm HG}$ acceleration is consistent with the predicted signal induced by 51~Eri~b. 

Recently, \citet{Brandt:2018dj} investigated potential systematic offsets between the {\it Hipparcos} and {\it Gaia} astrometric measurements and used a linear combination of the two {\it Hipparcos} reductions in an attempt to reduce observed discrepancies between the two catalogues. The revised proper motions presented within this catalogue are not significantly different for 51~Eri (Figure~\ref{fig:accl1}, red symbols), and is virtually unchanged for the most discrepant of the three accelerations ($\mu_{\rm G}-\mu_{\rm HG}$). 

\begin{figure}
\includegraphics[width=1.0\columnwidth]{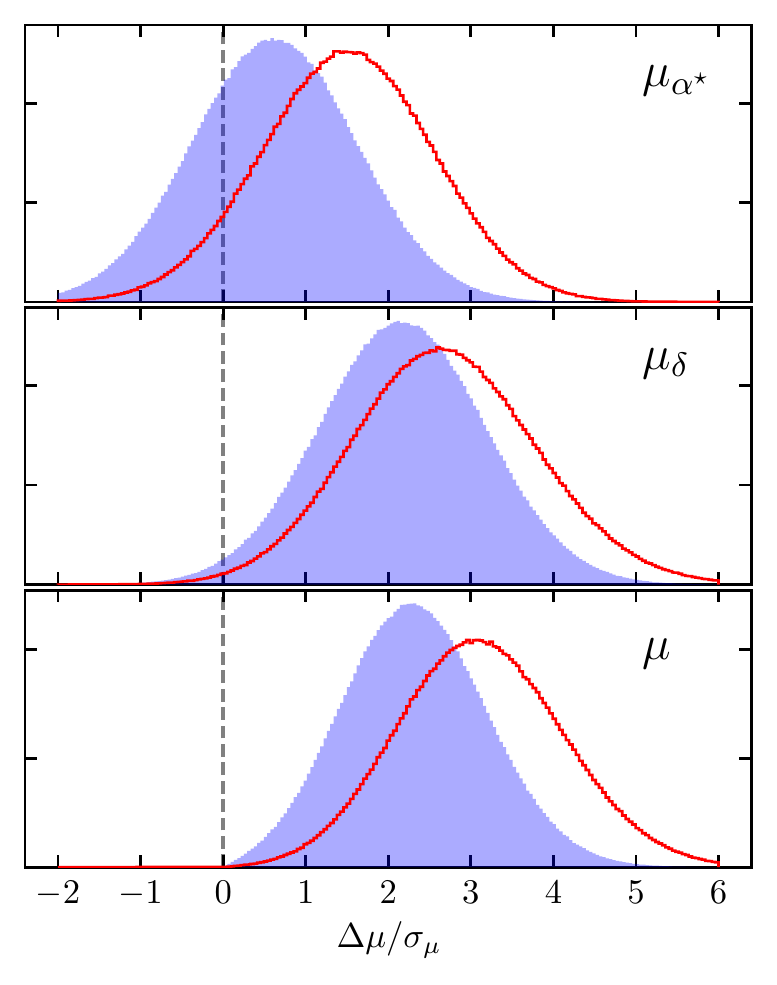}
\caption{Distribution of the difference between predicted and measured $\mu_{\rm G}-\mu_{\rm HG}$ divided by the uncertainty on the measurement for low-mass (2\,M$_{\rm Jup}$, blue filled histograms) and high-mass (12\,M$_{\rm Jup}$, red open histograms) planets in the right ascension (top panel) and declination (middle panel) directions, and the total magnitude of the acceleration (bottom panel) assuming symmetric uncertainties on the measured acceleration.
\label{fig:pm_significance}}
\end{figure}
\begin{figure*}
\includegraphics[width=1.0\textwidth]{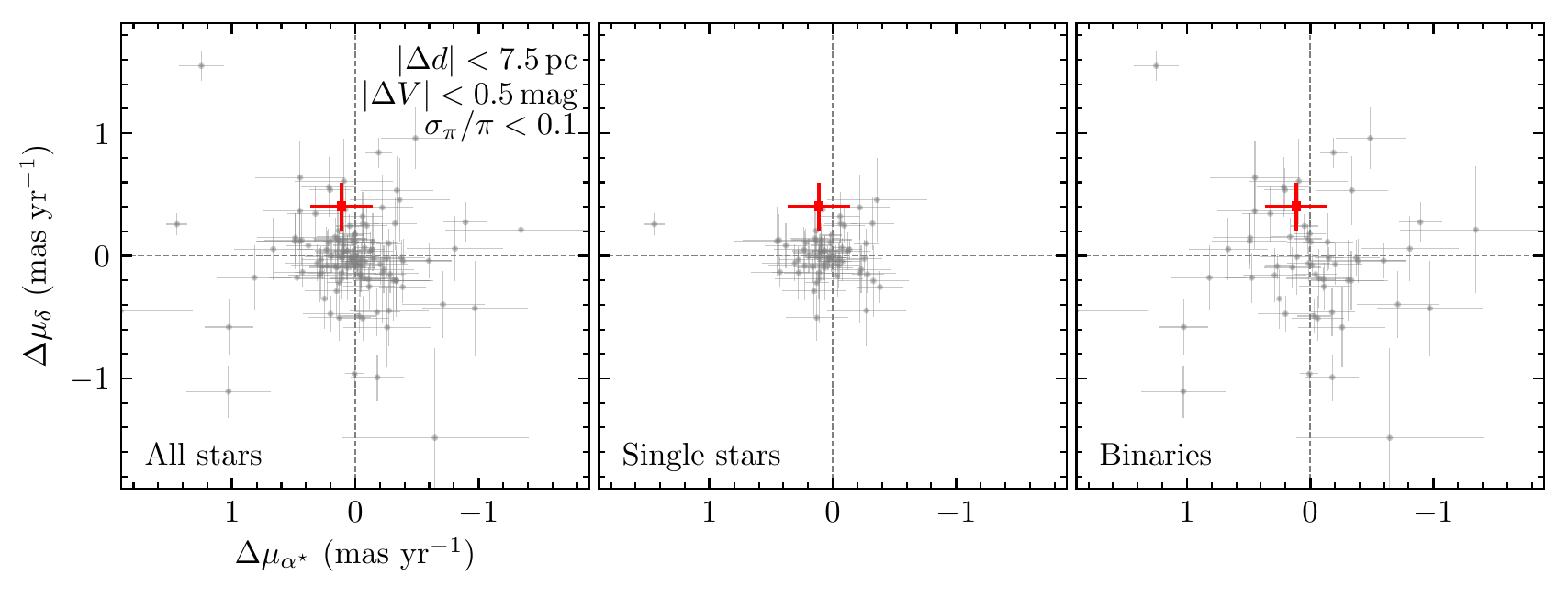}
\caption{Astrometric acceleration measured between the {\it Gaia} and the long-term proper motion for a sample of 155 stars (gray points) that share a similar distance, magnitude, and parallax uncertainty to 51 Eri (red). The three panels show the full sample (left), the 71 single stars (middle), and the 84 stars with evidence of binarity in the literature (right). \label{fig:pm_comparison}}
\end{figure*}

Figure~\ref{fig:pm_significance} shows the significance of the difference between the measured acceleration computed from the {\it Gaia} and long-term proper motions ($\mu_{\rm G}-\mu_{\rm HG}$), and that predicted from the visual orbit fit given in Section~\ref{sec:orbits}. The predicted acceleration from this combination of proper motions is the most constrained due to the relative astrometric record covering the same baseline as the {\it Gaia} mission. If we assume a mass of 2\,M$_{\rm Jup}$ for 51~Eri~b, the measured acceleration is $2.3\sigma$ discrepant ($0.6\sigma$ and $2.2\sigma$ in the $\alpha^{\star}$ and $\delta$ directions), rising to $3.1\sigma$ ($2.7\sigma$ and $1.5\sigma$) for a 12\,M$_{\rm Jup}$ planet.

\begin{figure}
\includegraphics[width=1.0\columnwidth]{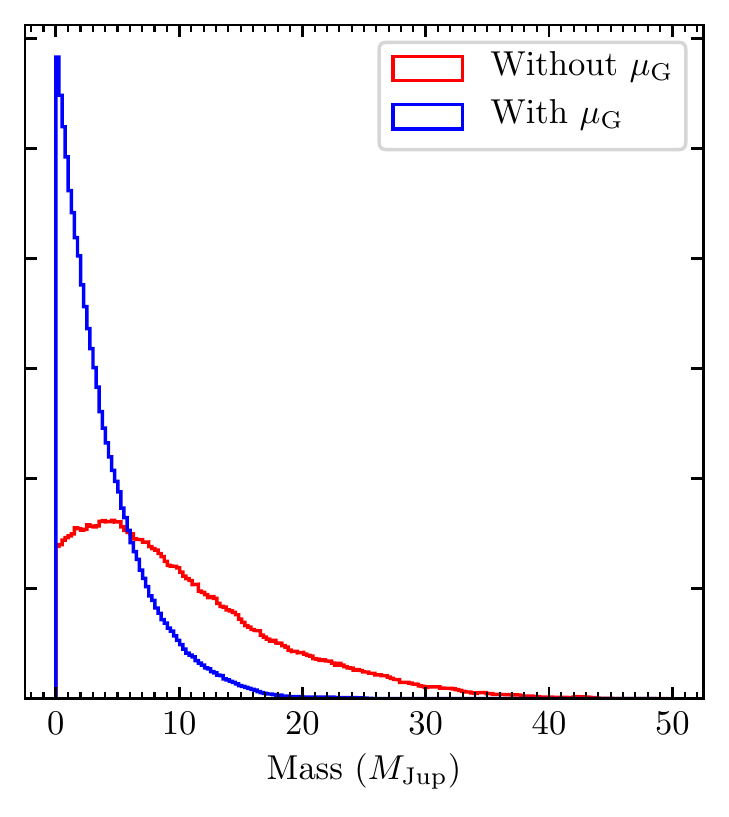}
\caption{Posterior distributions for the mass of 51~Eri~b from a joint fit to the relative and absolute astrometry of the system with (blue histogram) and without (red histogram) the {\it Gaia} proper motion. \label{fig:iad_m2}}
\end{figure}

The source of the discrepancy is not immediately apparent. 51~Eri ($G=5.1$) is close to the nominal bright limit of the astrometric instrument ($G=5$) when operating at the shortest integration times. The precision of the individual scan measurements at these magnitudes, between 1--2\,mas along the scan direction, is 25--50 times worse than the formal Poissonian uncertainties \citep{Lindegren:2018gy}. This difference was attributed primarily to inadequacies of the calibration models used to measure the centroid position of bright stars within each scan. It is not clear if these unmodelled errors would cause the centroid determination to be biased, or if they would simply introduce a random scatter on the measurement. It is plausible that the observed discrepancy is simply a random measurement error. This is more likely to be the case for a low-mass for 51~Eri~b, where the measurement is only a $2.3\sigma$ (roughly one-in-forty) outlier. If it is a measurement error, we are unable to differentiate between the low-mass and high-mass scenario for the planet at a significant level due to the marginal difference in the distributions shown in Figure~\ref{fig:pm_significance}. The high-mass scenario is approximately thirteen times less likely than the low-mass scenario (consistent with the relative probabilities in the mass posterior shown in Figure~\ref{fig:iad_m2}), and cannot be excluded at a significant level with the available measurements.

The discrepancy could also be astrophysical in nature. An additional companion to 51 Eri interior to the current sensitivity limits of instruments such as GPI and SPHERE in an appropriate orbit could be inducing the observed astrometric signal, either entirely or in combination with 51~Eri~b. As noted by \citet{Maire:2019vw}, a high eccentricity for the orbit of the planet could be the result of dynamical interactions with an additional companion within the system. To determine whether an astrophysical origin was a plausible source of the signal we compared the measured $\mu_{\rm G}-\mu_{\rm HG}$ acceleration for 51~Eri to a sample of 155 stars at a similar distance ($|\Delta d| < 7.5$\,pc), $V$-band magnitude ($|\Delta V| < 0.5$\,mag), and {\it Hipparcos} parallax uncertainty ($\sigma_{\pi}/\pi < 0.1$). We found 51~Eri to be a $1.2\sigma$ outlier when comparing to all stars in the sample (Figure~\ref{fig:pm_comparison}). However, the tails of the $\mu_{\rm G}-\mu_{\rm HG}$ distribution are undoubtedly contaminated with astrometric accelerations induced by stellar, substellar, and degenerate companions around these stars. We searched the Washington Double Star Catalog \citep{Mason:2001ff} to exclude binaries with a separation within $2\arcsec$, the Ninth Catalogue of Spectroscopic Binary Orbits \citep{Pourbaix:2004dg} to exclude spectroscopic binaries that can lead to spurious astrometric accelerations, and the Bright Star Catalogue \citep{Hoffleit:1991vr} for stars that had been categorized as being either variable radial velocity or a spectroscopic binary. We found evidence of binarity for 84 of the stars in the sample. Removing these binaries suppressed the tails of the distribution of astrometric accelerations for the 71 stars that to the best of our knowledge are single. The measured $\mu_{\rm G}-\mu_{\rm HG}$ acceleration for 51~Eri is more discrepant with this single subsample, a $1.5\sigma$ outlier, whereas it is consistent with the binary subsample. It is worth noting that not all of the stars within the single subsample have been searched for binary companions with either high-contrast imaging, interferometric observations, or radial velocity monitoring. The remaining outliers within this subsample are likely due to a combination of random measurement errors, systematic errors, and astrophysical signals induced by undiscovered companions.

These discrepancies have implications for an attempted measurement of the dynamical mass of 51~Eri~b with a joint fit to the visual orbit of the companion and the absolute astrometry of the host star. Using the framework described in Nielsen et al. (2019b, {\it submitted}), we performed two fits to the available data. The first used all available astrometry of the planet and host star, and the second excluded the {\it Gaia} proper motion due to the observed discrepancy in Figure~\ref{fig:accl1}. Both fits utilized the {\it Hipparcos} IAD rather than the {\it Hipparcos} catalogue values given in Table~\ref{tbl:hip_gaia}. The fit including the {\it Gaia} proper motion leads to a 1$\sigma$ upper limit on the planet mass of $M_2 < 7$\,M$_{\rm Jup}$, compared to $M_2 < 18$\,M$_{\rm Jup}$ from the fit where it is excluded. Based on the discrepancy between the predicted and measured value of $\mu_{\rm G}-\mu_{\rm HG}$ (and to a lesser extent $\mu_{\rm H}-\mu_{\rm G}$), we cannot use the former mass constraint to confidently rule out a high mass, low entropy formation scenario for 51~Eri~b. Instead, it is plausible that the fit is being driven towards to the lowest masses in an attempt to minimize the the $\mu_{\rm G}-\mu_{\rm HG}$ signal induced by the planet which is in the opposite direction to the measurement. A similar discrepancy between the predicted and measured {\it Gaia} proper motions is seen for $\beta$~Pic~b (Nielsen et al. 2019b, {\it submitted}), and was not used to constrain the mass of that planet.

\section{Future Mass Constraints with Gaia}
\label{sec:gaia}
The analyses presented in the previous sections are based on a comparison of the {\it Hipparcos} and {\it Gaia} catalogue proper motions. These measurements represent the combination of $\sim10^2$ individual astrometric measurements from each mission, fit based on an assumption of linear motion of the photocenter of the system. With sufficient astrometric precision, the reflex motion of the photocenter induced by the planet can be detected and used in conjunction with the visible orbit to constrain the mass of the planet. While the precision of the individual {\it Hipparcos} scan measurements ($\sigma=1.0\pm0.3$\,mas) is not sufficient to measure the expected displacement of the photocenter over the 2.5-year mission, the formal scan uncertainties for the final {\it Gaia} catalogue are predicted to be significantly lower.

\begin{figure}
\includegraphics[width=1.0\columnwidth]{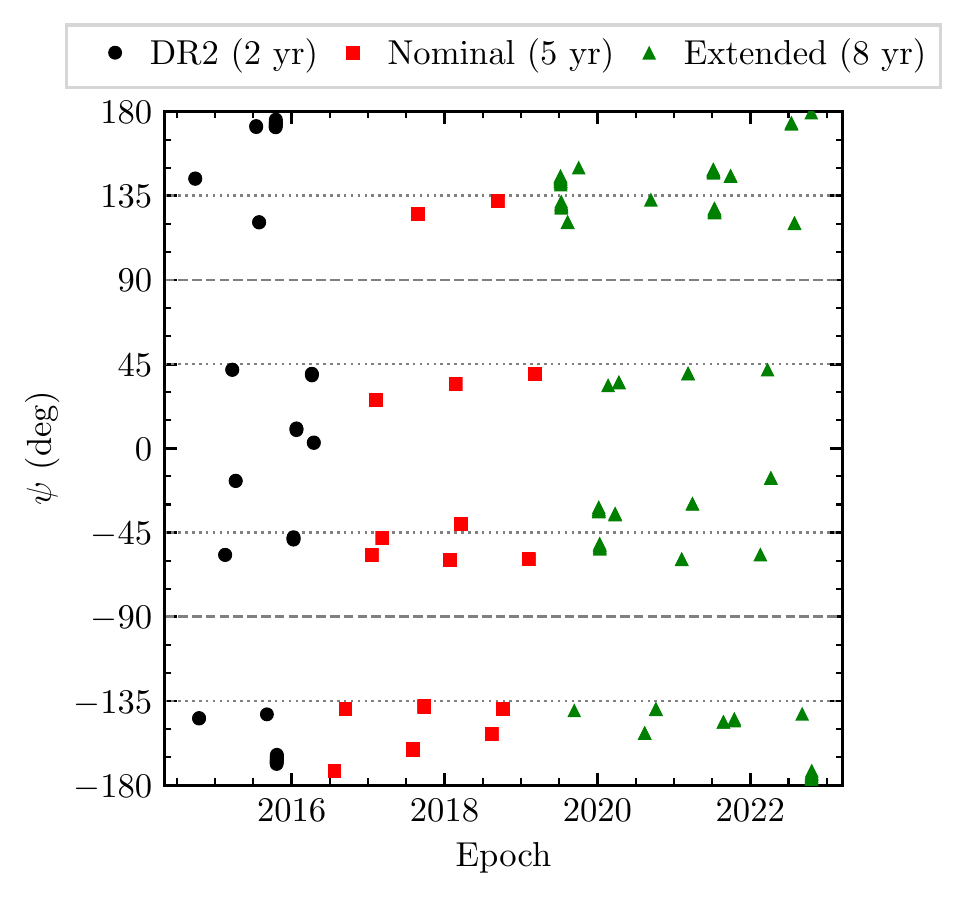}
\caption{Scan angles for the {\it Gaia} measurements of 51~Eri~b over the two years used to construct the DR2 catalogue (black circles), the nominal five-year mission (red squares), and an extended eight-year mission (green triangles). A scan angle of $|\psi|=90$\,deg (dashed) corresponds to a scan along the right ascension direction, constraining the position of the star only in that direction. Scan angles of $|\psi|=45$ and 135\,deg (dotted) provide equal constraints on the position of the star along the two axes. \label{fig:gaia_scans}}
\end{figure}
We utilized a similar framework to the one described in Section~\ref{sec:pred_accl} to assess the potential of {\it Gaia} observations alone to constrain the mass of the planet. For the purposes of these simulations we assumed that there are no additional massive companion within the system. We simulated a set of {\it Gaia} scan measurements of the 51~Eri system spanning three baselines, from the start of the mission (2014 July 25) to the end of the DR2 phase (2016 May 23), the end of the nominal five-year mission (assumed to be 2019-03-09), and the end of an extended eight-year mission (assumed to be 2022-12-31).

\begin{figure}
\includegraphics[width=1.0\columnwidth]{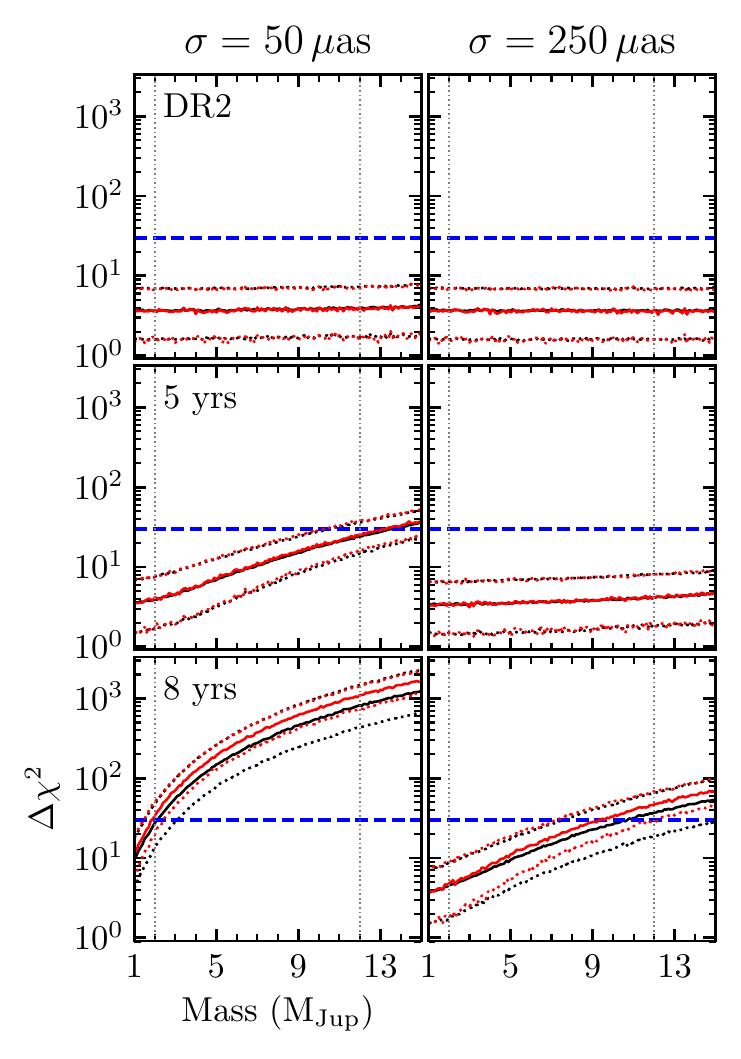}
\caption{Median (sold line) and 1$\sigma$ range (dotted line) of the $\Delta\chi^2$ between the five and twelve-parameter fit of the simulated {\it Gaia} abscissa measurements as a function of planet mass assuming a per-scan uncertainty of 50\,$\mu$as (left column) and 250\,$\mu$as (right column). The fits were performed on simulated measurements spanning the DR2 epoch (top row), the nominal five-year mission (middle row), and the extended eight-year mission (bottom row). The black curves are for the full set of visual orbits, and the red curves are for a subset of orbits consistent with a simulated epoch of relative astrometry in 2021.9. The model-dependent mass of the planet lies between 2--12\,M$_{\rm Jup}$ (gray dotted lines), and the criteria for detection of $\Delta\chi^2 \ge 30$ is also shown (blue dashed line). \label{fig:gaia_sim}}
\end{figure}
Simulated abscissa measurements were generated by combining the linear motion of the 51~Eri barycenter with the orbital motion of the photocenter for each of the samples within the MCMC chains from Section~\ref{sec:orbits}. As with the model in Section~\ref{sec:pred_accl}, we assumed a nominal set of astrometric parameters for the system barycenter. Gaussian noise was added to the simulated measurements with an amplitude of either 50\,$\mu$as, corresponding to the predicted noise floor for fifth magnitude stars, or 250\,$\mu$as, intermediate to this and the current median uncertainty of the individual scan measurements \citep{Lindegren:2018gy}. Our noise model assumed all of the measurements were uncorrelated. To assess whether the astrometric signal induced by the orbit of 51~Eri~b would have been detected within each of these simulation we fit the data with (1) a five-parameter model describing only the apparent motion of the system barycenter and (2) a twelve-parameter model that also accounts for the motion of the photocenter due to the orbiting planet. The first fit is performed as described in Section~\ref{sec:pred_accl}. For the second fit we used the framework described in \citet{Perryman:2014jr}. To speed up the optimization algorithm we fix the period of the planet and only fit the two non-linear terms $u$ and $v$ (transformed variables of the eccentricity $e$, and mean anomaly at the reference epoch $M_0$); the linear terms are determined exactly for each ($u$, $v$) pair. We computed the $\chi^2$ of each fit and consider the planet detected when $\Delta\chi^2 \ge 30$ \citep{Perryman:2014jr}.

The distributions of $\Delta\chi^2$ as a function of the mass of the planet are shown in Figure~\ref{fig:gaia_sim} for the two noise models. We find that the astrometric signal induced by the planet is only detectable ($>98$\% probability) in the simulations with the more favourable noise model (50\,$\mu$as scan uncertainty), with planet masses of $M_2\gtrsim4$\,M$_{\rm Jup}$, and that use the full dataset from the extended eight-year mission. The only possibility of an astrometric detection of 51~Eri~b in the nominal five-year mission is if it was a 12\,M$_{\rm Jup}$ planet in a favourable orbital configuration, the highest mass predicted for the planet from the ``cold start'' low-accretion formation scenario \citep{Rajan:2017hq}. We predict the astrometric signal of the planet will not be detectable at any plausible mass assuming a per-scan uncertainty of 250\,$\mu$as, which is already a factor of 4--8 improvement upon the estimate of the per-scan scan uncertainty of the astrometry used to create the {\it Gaia} DR2 catalogue \citep{Lindegren:2018gy}.

We also predicted the effect of an additional epoch of relative astrometry on the detectability of an astrometric acceleration with {\it Gaia}. We simulated one epoch of astrometry in 2021.9 consistent with an eccentricity at the median of the marginalized distribution ($e=0.53$; $\rho=357.1\pm3.0$\,mas, $\theta=129\fdg1\pm0\fdg3$) and used rejection sampling to select the orbits consistent with this measurement. The $\Delta\chi^2$ distribution for this subset of orbits is not significantly different; the planet is not detectable except in the most favourable circumstances. Repeating this analysis for a simulated measurement consistent with a low ($e=0.40$) or high ($e=0.62$) orbital eccentricity did not lead to a significant change in the distribution of $\Delta\chi^2$ as a function of planet mass.

\section{Conclusion}
We have presented an update to the visual orbit of the young, low-mass directly imaged exoplanet 51 Eridani b using astrometry obtained with Gemini/GPI over the previous three years. We find orbital elements that are consistent with an independent analysis of a dataset combining literature GPI astrometry with new VLT/SPHERE measurements \citep{Maire:2019vw}, and within the uncertainties presented in an earlier analysis with a nine-month baseline by \citet{DeRosa:2015jl}. We can confidently exclude a highly eccentric orbit for the planet, but a degeneracy exists between inclined low-eccentricity ($e\sim0.2$) orbits and less inclined but more eccentric ($e\sim0.5$) orbits. This degeneracy can be broken with either long-term astrometric monitoring of the visual orbit, or in short order with a radial velocity measurement of the planet with instruments that combine high-contrast imaging techniques with high-resolution spectroscopy (e.g., \citealp{Wang:2017kv}). Previous radial velocity measurements for short-period directly-imaged exoplanets have used more traditional slit spectroscopy \citep{Snellen:2014kz}, a technique that is challenging for 51 Eri b given the high contrast between the planet and its host star.

With a revised visual orbit for the system, we predicted the astrometric signal induced on 51~Eri by the orbiting planet and compared to absolute astrometry from the {\it Hipparcos} and {\it Gaia} catalogues. We find that the predicted acceleration for the star due to the planet is inconsistent with the measured value at the 2--3$\sigma$ level and for one combination of catalogue proper motions the acceleration vector is in the opposite direction to that predicted by the visual orbit. This discrepancy could be due a combination of random measurement errors and other sources of uncertainty in the {\it Gaia} astrometry that have not been correctly modelled for bright stars \citep{Lindegren:2018gy}, or a real astrophysical signal induced by an additional companion within the system that is interior to current detection limits. This discrepancy precludes a dynamical mass determination or constraint using the currently available data. Finally, we performed simulations of the individual {\it Gaia} scan measurements of 51~Eri over the course of the extended eight-year {\it Gaia} mission. We demonstrated that a dynamical mass measurement of 51~Eri~b using {\it Gaia} data alone is only possible at $>98$\% confidence assuming the most optimistic predictions for the final per-scan uncertainty of the {\it Gaia} astrometry and a mass of $\gtrsim4$\,M$_{\rm Jup}$ for the planet.

The upcoming {\it Gaia} data releases will contain astrometric accelerations, photometric orbit fits, and the individual scan measurements used to construct the catalogue. Combined with long-term proper motions derived from {\it Hipparcos} positions (e.g., \citealp{Brandt:2018dj, Kervella:2019bw}), this rich resource will enable targeted searches for substellar companions to nearby, young stars that are amenable to direct detection, spectroscopic characterization, and eventual dynamical mass measurements. The release of this catalogue will be timely for the launch of the {\it James Webb Space Telescope}; the sensitivity of the thermal-infrared coroangraphic instruments will be sufficient to detect wide-orbit Jovians around much older (and typically closer) stars than have previously been targeted from the ground.

\acknowledgments
We are grateful to the referee who helped to improve the quality of this work. We thank Trent Dupuy for useful discussions relating to this work. Supported by NSF grants AST-1411868 (R.D.R., E.L.N., K.B.F., B.M., J.P., and J.H.), AST-141378 (G.D.), AST-1518332 (R.D.R., J.J.W., T.M.E., J.R.G., P.G.K.). Supported by NASA grants NNX14AJ80G (R.D.R., E.L.N., B.M., F.M., and M.P.), NSSC17K0535 (R.D.R., E.L.N., B.M., J.B.R.), NNX15AC89G and NNX15AD95G (R.D.R., B.M., J.E.W., T.M.E., G.D., J.R.G., P.G.K.). This work benefited from NASA's Nexus for Exoplanet System Science (NExSS) research coordination network sponsored by NASA's Science Mission Directorate. J.R is supported by the French National Research Agency in the framework of the Investissements d’Avenir program (ANR-15-IDEX-02), through the funding of the ``Origin of Life'' project of the University Grenoble-Alpes. Portions of this work were performed under the auspices of the U.S. Department of Energy by Lawrence Livermore National Laboratory under Contract DE-AC52-07NA27344. V.P.B. acknowledges government sponsorship. Portions of this work were carried out at the Jet Propulsion Laboratory, California Institute of Technology, under a contract with the National Aeronautics and Space Administration. J.J.W. is supported by the Heising-Simons Foundation 51~Pegasi~b postdoctoral fellowship. Based on observations obtained at the Gemini Observatory, which is operated by the Association of Universities for Research in Astronomy, Inc., under a cooperative agreement with the NSF on behalf of the Gemini partnership: the National Science Foundation (United States), National Research Council (Canada), CONICYT (Chile), Ministerio de Ciencia, Tecnolog\'{i}a e Innovaci\'{o}n Productiva (Argentina), Minist\'{e}rio da Ci\^{e}ncia, Tecnologia e Inova\c{c}\~{a}o (Brazil), and Korea Astronomy and Space Science Institute (Republic of Korea). This research used resources of the National Energy Research Scientific Computing Center, a DOE Office of Science User Facility supported by the Office of Science of the U.S. Department of Energy under Contract No. DE-AC02-05CH11231. This work used the Extreme Science and Engineering Discovery Environment (XSEDE), which is supported by National Science Foundation grant number ACI-1548562. This work has made use of data from the European Space Agency (ESA) mission {\it Gaia} (\url{https://www.cosmos.esa.int/gaia}), processed by the {\it Gaia} Data Processing and Analysis Consortium (DPAC, \url{https://www.cosmos.esa.int/web/gaia/dpac/consortium}). Funding for the DPAC has been provided by national institutions, in particular the institutions participating in the {\it Gaia} Multilateral Agreement. This research has made use of the SIMBAD database and the VizieR catalog access tool, both operated at the CDS, Strasbourg, France. This research has made use of the Washington Double Star Catalog maintained at the U.S. Naval Observatory.

\facility{Gemini:South (GPI)}

\software{Astropy \citep{TheAstropyCollaboration:2013cd},  
          Matplotlib \citep{Hunter:2007ih},
          pyKLIP \citep{Wang:2015th}}

\bibliographystyle{aasjournal}   
\bibliography{refs}

\end{document}